\documentclass[11pt]{article}
\begin{document}
\title
{Interaction of a circularly polarised gravitational wave with a charged particle in a static magnetic background}
\author{
{\bf {\normalsize Sunandan Gangopadhyay}$^{a,b}
$\thanks{sunandan.gangopadhyay@gmail.com}},
{\bf {\normalsize Anirban Saha}
$^{a,b}$\thanks{anirban@iucaa.ernet.in}},
{\bf {\normalsize Swarup Saha}$^{a,c}$\thanks{saha18swarup@gmail.com}}\\
$^{a}$ {\normalsize Department of Physics, West Bengal State University, Barasat, India}\\
$^{c}$ {\normalsize Department of Radiotherapy and Nuclear Medicine , Barasat Cancer Research and Welfare Center}\\
{\normalsize Barasat, India}\\[0.3cm]
$^{b}${\normalsize Visiting Associate in Inter University Centre for Astronomy $\&$ Astrophysics,}\\
{\normalsize Pune, India}\\[0.3cm]
}
\date{}
\maketitle
\begin{abstract}
{\noindent Interaction of a charged particle in a static magnetic background, i.e., a Landau system with circularly polarised gravitational wave (GW) is studied quantum mechanically in the long wavelength and low velocity limit. We quantize the classical Hamiltonian following \cite{speli}. The rotating polarization vectors of the circularly polarized GW are employed to form a unique directional triad which served as the coordinate axes. The Schrodinger equations for the system are cast in the form of a set of coupled linear differential equations. This system is solved by iterative technique. We compute the time-evolution of the position and momentum expectation values of the particle. The results show that the resonance behaviour obtained earlier\cite{emgw_classical} by classical treatements of the system has a quantum analogue not only for the linearly polarized GW \cite{emgw_1_lin}, but for circularly polarized GW as well.}
\end{abstract}
\maketitle

  With the development of various ground based gravitational wave (GW) detectors like LIGO \cite{abramovici}, the possibility of direct detection of GW(s) with a strain sensitivity of the order of $h \sim 10^{-21}/\sqrt{Hz} $ or better in the frequency range between 100-1000 Hz is expected in near future \cite{thorn}. With such a small response of matter to a passing GW, more often than not, a quantum mechanical treatment of the matter-GW interaction is desirable \cite{caves, weber}. Moreover, recent phenomenologicl studies of NCQM \cite{pmh, bert0, ani, sgfgs}  and NC quantum field theory (NCQFT) \cite{carol, stern, camel} had estimated the upper-bounds of the noncommutative length-scale to be of the same order as the displacement of a test particle under GW. Thus a good prospect of identifying the NC nature of spacetime (or at least putting stringent upper-bounds on the NC parameters \cite{bert0, ani, carol, stern, cst, RB}) requires modeling the GW detection experiments in an NC setting and confront the model with GW detection data. Constructing such NC models essentially require a priori quantum mechanical setting.
The recent studies in commutative quantum mechanics \cite{speli} ans noncommutative quantum mechanics (NCQM) \cite{ncgw1, ncgw2, ncgw_1_circular} of harmonic oscillators interacting with GW show interesting resonance features in the result which can be useful to probe the noncommutative structure of space in GW detector read-outs.
A natural realization of such quantum harmonic oscillators is the Landau system where a charged particle in a static magnetic background is studied quantum mechanically. So it will be interesting to investigate the quantum mechanics of the Landau system under the influence of GW. Once such a quantum mechanical analysis is available it can be readily elevated to NC level using the established framework of NC electrodynamics \cite{sch, bcsgas, bcsgfgs} and NC gravity \cite{grav, sgrab}.

Although some investigations on the coupling of a background magnetic field with GW at a classical level has been carried out in the literature \cite{emgw_classical}, a quantum mechanical treatement of the same has been missing and is of foremost interest in its own right.  Such a quantum mechanical formulation would allow us to compare our results with the classical results and give us deeper insight. Furthermore, in recent years there is a resurgence of interest in this age-old problem of the coupling of electromagnetic field to GW since experimental investigations are under way with ever increasing precision both in the laboratory or by astronomical observation \cite{Tsagas, Marklund}. For example, analytic treatement and numerical simulation of the interaction of a GW with a strongly magnetized plasma have shown that for strong magnetic fields ($\sim 10^{15} {\rm Gauss}$), the GW excites electromagnetic plasma waves where the energy absorbed from the GW by the electromagnetic oscillations is comparable to the energies emitted in the most energetic astrophysical events, such as giant flares on magnetars and possibly even short gamma ray bursts \cite{Isliker}. Evidently, quantum mechanical models of such astrophysical phenomena, where coupled effect of GW and electromagnetic field set the stage for charged matter to evolve, can be very interesting. Towards this endeavour, we first have to consider the response of a single charge to both electromagnetic field and GW simultaneously at the quantum mechanical level. Also, since GW couples to electromagnetic waves in plasma mainly through the generation of electric currents inside the plasma by perturbing the charged particle trajectories \cite{Papa}, the quantum mechanics of the charged particle influenced by GW will also be relevant in that context. 


With this motivation we have studied the effect of GW on the Landau system in an earlier paper \cite{emgw_1_lin} where for simplicity we have focused solely on the linearly polarized GW and left out an important category of GWs, namely the circularly polarized ones. However, the circular polarization of GW gives a way to describe whether the background has asymmetry with respect to magnitudes of right-handed and left-handed GWs \cite{kah, seto}. Interestingly, the most recent BICEP results \cite{BICEP} claim to have detected the B-type polarization mode in the cosmic microwave background (CMB) which is caused by the left/right circular polarized GWs. This detection, though indirect in nature, has reinforced the relevance of circularly polarized GWs. Thus it is imparative that we extend our earlier work \cite{emgw_1_lin} for the circularly polarized GW.

In the present paper we therefore study the Landau system interacting with a circularly polarized GW. Note that due to rotation of the GW polarization vectors in the present case, the method of analysis presented in \cite{emgw_1_lin} will have to be modified in a non-trivial way which will be discussed shortly. To specify, we shall study the quantum mechanics of a charged particle gyrating in the presence of a constant background magnetic field along the z-axis, while circularly polarized GW(s) parallel to the magnetic field is passing. The classical approach to this problem is to solve the geodesic equation of a charged particle in a uniform magnetic filed in a linearized gravity background. However we choose the geodesic deviation equation for the system as our starting point and work backwords to obtain the Hamiltonian describing the system. This Hamiltonian is eventually quantized \cite{speli} to go over to the quantum mechanical scenario. The classical treatment of this system where the GW, with a frequency $\Omega$, propagates parallel to the magnetic field, the coupling of a gyrating charge with the GW becomes very strong at twice the cyclotron frequency $\Omega = 2\varpi$ \cite{emgw_classical}. In \cite{emgw_1_lin} we have shown that for linearly polarized GW the quantum mechanical model also shows such resonance. To check whether this behaviour persists for circularly polarized GW is one of our main objectives in the present paper.

The argument of working back from the geodesic deviation equation in the proper detector frame to the corresponding Hamiltonian has been presented in details in \cite{ncgw1, ncgw2}. Following the same path, the Hamiltonian in the present case becomes 
\begin{equation}
{H} = \frac{1}{2m}\left({p}_{j} - \frac{e}{c}A_{j} + m \Gamma^j_{0k} {x}_{k}\right)^2  
\label{e10}
\end{equation}
where standard notations aer used. 

Owing to the transverse nature, for the GW propagating along the $z$-axis, the objects ${\Gamma^j}_{0k}$ will be confined only in the ${x}-{y}$ plane. We employ the symmetric gauge $A_{j} = -\frac{B}{2} \epsilon_{jl}x_{l}$ in that plane and using the traceless property of the GW, the Hamiltonian (\ref{e10}) takes the form
\begin{equation}
{H} = \frac{1}{2m}{p}^{2}_{j} + \frac{1}{2} m \varpi^{2}x^{2}_{l} -\varpi L  + \frac{1}{2}\dot{h}_{jk} x_{k}p
_{j} - \frac{eB}{4 c} \epsilon_{lj} \dot{h}_{jk} x_{l} {x}_{k} ~. 
\label{e11}
\end{equation}
As we have pointed out earlier the first two terms represent an ordinary harmonic oscillator with the cyclotron frequency $\varpi = \frac{e B}{2m c}$. The third one is the Zeeman term with $L = \epsilon_{lj} x_{l} p_{j}$ and the fourth term which is linear in the affine connections, reveals the effect of the passing GW. Finally, the last term presents the coupling between the GW and the background magnetic field. Since we are dealing with linearized gravity, a term quadratic in $\Gamma$ has been neglected in eq.(\ref{e11}). Thus we have arrived at a Hamiltonian conisting of a harmonic oscillator along with other smaller terms that can be regarded as perturbations. 

We define raising and lowering operators in terms of the oscillator frequency $\varpi$ as
\begin{eqnarray}
x_j &=& \left({\hbar\over 2m\varpi}\right)^{1/2}
\left(a_j+a_j^\dagger\right)\>\label{e15a} \\
p_j &=& -i\left({\hbar m\varpi\over 2}\right)^{1/2} 
\left(a_j-a_j^\dagger\right)\>
\label{e15}
\end{eqnarray}
to write the Hamiltonian (\ref{e11}) as 
\begin{eqnarray}
{\hat H} &=& \hbar\varpi\left( a_j^\dagger a_j + 1 \right) + i \hbar\varpi \epsilon_{ij} a^{\dagger}_{i}a_{j} - \frac{i\hbar}{4} \dot h_{jk} 
\left(a_j a_k - a_j^\dagger a_k^\dagger\right)  \nonumber \\
&& - \frac{\hbar}{4} \left[\epsilon_{lj} {\dot h}_{jk}  \left(a_{l}a_{k}  - a_{l}^{\dagger} a_{k}^\dagger \right) 
+
 \left(\epsilon_{lj} {\dot h}_{jk} + \epsilon_{kj} {\dot h}_{jl}\right) a_{l}^{\dagger} a_{k}\right]
\label{e16}
\end{eqnarray}
so that the equation of motion for $a_{i}(t)$ will be 
\begin{eqnarray}
\frac{da_{i}(t)}{dt} &=& -i{\varpi}a_i + \varpi \epsilon_{ij}a_{j} + \frac{1}{2}\dot h_{ij}a^\dagger_j 
+ \frac{i}{4} \epsilon_{ij} \dot h_{jk} \left(a^{\dagger}_{k} + a_{k}\right) + \frac{i}{4} \epsilon_{kj} \dot h_{ji} \left(a^{\dagger}_{k} + a_{k}\right).
\label{e17}
\end{eqnarray}
To obtain the same for of $a_{i}^{\dagger}(t)$ one has to take the complex conjugate (c.c) of the above equation. Since the raising and lowering operators satisfy the commutation relations
\begin{eqnarray}
\left[a_j(t), a^\dagger_k(t)\right] = \delta_{jk}, \qquad
\left[a_j(t), a_k(t)\right] = 0 = 
\left[a^\dagger_j(t), a^\dagger_k(t)\right]
\label{e18}
\end{eqnarray}
we can study their time-evolutions using the time-dependent Bogoliubov transformations. To this end we express the raising and lowering operators in terms of the free operators at time $t=0$ and the generalised Bogoliubov coefficients  $u_{jk}$ and $v_{jk}$
\begin{eqnarray}
a_j(t) &=& u_{jk}(t) a_k(0) + v_{jk}(t)a^\dagger_k(0)\> \nonumber \\
a_j^\dagger(t) &=& a_k^\dagger(0)\bar u_{kj}(t)  + a_k(0)\bar v_{kj}(t).\>
\label{e19}
\end{eqnarray}
Here the bar denotes the complex conjugation (c.c). The  Bogoliubov coefficients are $2\times 2$ complex matrices which, owing to the commutation relations (\ref{e18}) must satisfy
$uv^{T}=u^{T}v\>,\> u u^\dagger - v v^\dagger = I,$
where $T$ denotes transpose, $\dagger$ denotes c.c transpose and $I$ is the identity
matrix. 
$u_{jk}(t)$ and $v_{jk}(t)$ satisfy the appropriate boundary conditions 
\begin{eqnarray}
u_{jk}(0)& = & I  \quad,\quad  v_{jk}(0) = 0~.
\label{bc0}
\end{eqnarray}
In terms of the variables $\zeta = u - v^\dagger$ and $\xi = u + v^\dagger$ the equations of motions eq.(\ref{e17}) and its c.c, take the form 
\begin{eqnarray}
\frac{d \xi_{il}}{dt} &=& -i\varpi \zeta_{il} + \varpi \epsilon_{ij} \xi_{jl} + \frac{1}{2}{\dot h}_{ij}\xi_{jl} \>
\label{e21a}\\
\frac{d \zeta_{il}}{dt} &=& -i\varpi \xi_{il} +  \varpi \epsilon_{ij} \zeta_{jl} - \frac{1}{2}{\dot h}_{ij}\zeta_{jl} 
+ \frac{i}{2} \epsilon_{ij} {\dot h}_{jk}\xi_{kl} - \frac{i}{2} {\dot h}_{ij} \epsilon_{jk} \xi_{kl} ~. 
\label{e21b} 
\end{eqnarray}
The eq(s)(\ref{e21a}, \ref{e21b}) are to be solved for the case of circularly polarized GW(s). 

\noindent In the two-dimensional plane, the circularly polarized GW, which is a $2\times 2$ matrix $h_{jk}$, can be written in terms of the first and third Pauli spin matrices $\sigma^A_{jk}, A=1,3$ and the two rotating polarization vectors $\varepsilon_{A}$ as 
\begin{equation}
h_{jk} \left(t\right) = 2f_{0}\left(\varepsilon_{1}(t)\sigma^1_{jk} + \varepsilon_{3}(t)\sigma^3_{jk}\right) 
= 2f_{0}~\varepsilon_{A}(t)\sigma^A_{jk}.
\label{e13}
\end{equation}
Here $2f_{0}$ is the constant amplitude of the GW and the polarization vectors $\varepsilon_{1}(t)$ and $\varepsilon_{3}(t)$  satisfy the constraint $\varepsilon_{1}^2+\varepsilon_{3}^2 = 1$ for all $t$ and evolve according to 
\begin{equation}
\frac{d\epsilon_{3}}{dt} = \Omega
\epsilon_{1}\>,\qquad\frac{d\epsilon_{1}}{dt} = - \Omega
\epsilon_{3}\>.
\label{e23}
\end{equation}
with a constant frequency $\Omega$.

To solve eqs.$(\ref{e21a})$ and $(\ref{e21b})$ let us first note that any $2\times 2$ complex matrix $M$ can be expressed as a linear combination of the Pauli spin matrices and identity matrix as
\begin{equation}
M = \theta_{0} I + \theta_{A} \sigma^A
\label{e23s}
\end{equation} 
where $\theta_0$ and $\theta_A$ are complex quantities. Note that $\vec\theta = (\theta_A), \, A=1,2,3$ can be thought of as a vector in a three dimensional complex space. The polarization states of the GW can also be represented as a vector $\vec\varepsilon$ in such a complex space. Also note that $\vec\varepsilon$, $\dot{\vec\varepsilon}$ and $\vec\varepsilon\times\dot{\vec\varepsilon}$ are mutually orthogonal and thus can serve as a natural choice of mutually independent directions, thus forming a coordinate system for our purpose. Hence, we make the following ansatz: 
\begin{equation}
\zeta = A I + B\vec\varepsilon\cdot\vec\sigma +
       C\frac{\dot{\vec\varepsilon}\cdot\vec\sigma}{\Omega} +
       Di\frac{\vec\varepsilon\times
       \dot{\vec\varepsilon}}{\Omega}\cdot\vec\sigma\>,
\label{e24}
\end{equation}
\begin{equation}
\xi =  E I + F\vec\varepsilon\cdot\vec\sigma +
       G\frac{\dot{\vec\varepsilon}\cdot\vec\sigma}{\Omega} +
       Hi\frac{\vec\varepsilon\times
       \dot{\vec\varepsilon}}{\Omega}\cdot\vec\sigma\>,
\label{e25}
\end{equation}
where $A$, $B$, $C$, $D$,$E$,$F$,$G$,$H$ can be complex functions. Substituting 
from eq(s)(\ref{e13}, \ref{e23}, \ref{e24},\ref{e25}) in eq(s)(\ref{e21a}, \ref{e21b}) and comparing the coefficients 
we get a set of first order differential equations for the complex functions $A, B, C ,D ,E ,F ,G ,H $:
\begin{eqnarray}
\frac{dA}{dt}+ i\varpi E -\varpi A +  f_0\Omega C+  2i f_0\Omega F &=& 0\>,
\nonumber \\
\frac{dB}{dt} - \Omega C+ i\varpi F-\varpi B -f_0\Omega D +  2i f_0\Omega E&=& 0\>,
\nonumber \\
\frac{dC}{dt} + \Omega B + i\varpi G-\varpi C + f_0\Omega A +  2i f_0\Omega H&=& 0\>,
\nonumber \\
\frac{dD}{dt} + i\varpi H -\varpi D - f_0\Omega B +  2i f_0\Omega G &=& 0\>
\nonumber \\
\frac{dE}{dt} +i\varpi A -\varpi E - f_0\Omega G  &=& 0\>,
\nonumber \\
\frac{dF}{dt} - \Omega G+i\varpi B-\varpi F  + f_0\Omega H  &=& 0\>,
\nonumber \\
\frac{dG}{dt} + \Omega F +i\varpi C -\varpi G - f_0\Omega E &=& 0\>,
\nonumber \\
\frac{dH}{dt} +i\varpi D -\varpi H + f_0\Omega F  &=& 0\>.
\end{eqnarray}
Solving them with boundary conditions appropriate with (\ref{bc0}), we obtain to first order in the GW amplitude, 
\begin{eqnarray}
A(t)&=&1+\frac{\Omega f_0 J_3}{2}-\frac{\left(1+J_2\right)+i\left(1+J_1\right)}{4}-i f_0\left[\Omega J_4 - \frac{2\varpi}{\Omega-2\varpi}\right]\nonumber\\
B(t)&=&-\frac{\Omega f_0\left(1+J_2\right)}{4\varpi}-\frac{\left(\Omega-\varpi \right)J_3}{2}-i\left[\frac{\varpi J_4}{2}+\frac{f_0\Omega\left(1+J_1\right)}{2\varpi}-\frac{\varpi^2}{\Omega\left(\Omega-2\varpi\right)}\right]\nonumber\\
C(t)&=&-\frac{\Omega f_0\left(1+J_1\right)}{4\varpi}-\frac{\left(\Omega-\varpi \right)J_4}{2}+\frac{\varpi\left(\Omega-\varpi\right)}{\Omega\left(\Omega-2\varpi\right)}-i\left[\frac{\varpi J_3}{2}-\frac{f_0\Omega\left(1+J_2\right)}{2\varpi}\right]\nonumber\\
D(t)&=&\frac{\Omega f_0 J_4}{2}-\frac{f_0\varpi}{\left(\Omega-2\varpi\right)}-\frac{\left(1+J_1\right)-i\left(1+J_2\right)}{4}-i f_0\Omega J_3\nonumber\\
E(t)&=&1-\frac{\Omega f_0 J_3}{2}-\frac{\left(1+J_2\right)+i\left(1+J_1\right)}{4}\nonumber\\
F(t)&=&\frac{\Omega f_0\left(1+J_2\right)}{4\varpi}-\frac{\left(\Omega-\varpi \right)J_3}{2}-i\left[\frac{\varpi J_4}{2}-\frac{\varpi^2}{\Omega\left(\Omega-2\varpi\right)}\right]\nonumber\\
G(t)&=&\frac{\Omega f_0\left(1+J_1\right)}{4\varpi}-\frac{\left(\Omega-\varpi \right)J_4}{2}+\frac{\varpi\left(\Omega-\varpi\right)}{\Omega\left(\Omega-2\varpi\right)}-i\frac{\varpi J_3}{2}\nonumber\\
H(t)&=&-\frac{\Omega f_0 J_4}{2}+\frac{f_0 \varpi}{\left(\Omega-2\varpi\right)}-\frac{\left(1+J_1\right)-i\left(1+J_2\right)}{4}\nonumber\\
\label{100y}
\end{eqnarray}
where
\begin{eqnarray}
J_1&=& 2\varpi t + \sin2\varpi t -\cos2 \varpi t \nonumber\\
J_2&=& 2\varpi t - \sin2\varpi t -\cos2 \varpi t \nonumber\\
J_3&=& \frac{\sin\left(\Omega-2\varpi\right)t}{\Omega-2\varpi}-\frac{\sin\Omega t}{\Omega}\nonumber\\
J_4&=& \frac{\cos\left(\Omega-2\varpi\right)t}{\Omega-2\varpi}-\frac{\cos\Omega t}{\Omega}
\label{100z}
\end{eqnarray}
With the above expressions for $A, B, C, D, E, F, G, H $ computed, the system has now been essentially solved. The time evolution of various relevant measurable quantities (e.g., coordinate and momentum expection values) can be obtained as follows. 

Combining the expressions for $A, B, C, D, E, F, G, H $ given by (\ref{100y}, \ref{100z}), we can write the solutions in terms of $\zeta$ and $\xi$ using Eq.(s) $(\ref{e24}, \ref{e25})$ which in turn give $u$ and $v$. Using $u$ and $v$ in Eq.(\ref{e19}) we can now compute the raising/lowering operators $a_j(t)$ and $a_j^\dagger(t)$ once their initial values are fixed. To fix the initial values of the raising/lowering operators, one only needs to specify the initial values (or more conveniently the initial expection values) of the position $\vec{r}_{0} = \left(x_{0}, y_{0}\right)$ and momentum $\vec{p}_{0} = \left(p_{x_{0}}, p_{y_{0}}\right)$ when the GW just hits the system (say,) at time $t=0$. From the initial position and momentum expectation values, i.e.$\langle\vec{r}_{0} \rangle= \left(X_{1}(0), X_{2}(0)\right)$ and $\langle\vec{P}_{0}\rangle = \left(P_{1}(0), P_{2}(0)\right)$, we get the corresponding values for the raising and lowering operator $\langle a_{j}\left( 0 \right)\rangle$ and $\langle a_{j}^{\dagger}\left( 0 \right)\rangle$, taking the expection values of equations (\ref{e15a}, \ref{e15}) at time $t = 0$.
We then use them in Eq.(s) $(\ref{e19})$ to find $a_{j}\left( t \right)$ and $a_{j}^{\dagger}\left( t \right)$ at a general time $t$ and these yield the time evolution of the expectation values of position coordinates $\langle X_{1} \left(t \right) \rangle$ and $\langle X_{2} \left(t \right)\rangle$ of the test body. Their general expressions, thus obtained, are given by 
\begin{eqnarray}
\langle X_{1}\left(t\right)\rangle & = & \left[\left(1- \frac{1+J_2}{4}\right)X_{1}\left( 0 \right)+\frac{\left(1+J_1\right)}{4}\frac{P_{1}\left( 0 \right) }{m\varpi}\right]+\left[\left(\frac{1+J_1}{4}\right)X_{2}\left( 0 \right)+\frac{\left(1+J_2\right)}{4}\frac{P_{2}\left( 0 \right) }{m\varpi}\right]\nonumber\\
&& + \frac{\Omega f_0}{2}\left[J_3X_{1}\left( 0 \right)-J_4X_{2}\left( 0 \right)\right]+\frac{\varpi\left(\Omega-\varpi\right)}{\Omega\left(\Omega-2\varpi\right)}\left[\epsilon_1X_{1}\left( 0 \right)-\epsilon_3X_{2}\left( 0 \right)\right] \nonumber\\
&& +\frac{\left(\Omega-\varpi\right)}{2}\left[-\left(\epsilon_3J_3+\epsilon_1J_4\right)X_{1}\left( 0 \right)+\left(\epsilon_3J_4-\epsilon_1J_3\right)X_{2}\left( 0 \right)\right]\nonumber\\
&&+\frac{\Omega f_0}{4\varpi}\left[-\left\{\left(1+J_2\right)\epsilon_3+\left(1+J_1\right)\epsilon_1\right\}X_{1}\left( 0 \right)+\left\{\left(1+J_1\right)\epsilon_3-\left(1+J_2\right)\epsilon_1\right\}X_{2}\left( 0 \right)\right]\nonumber\\
&&+\frac{1}{2m}\left[\left(\epsilon_3J_4-\epsilon_1J_3\right)P_{1}\left( 0 \right)+\left(\epsilon_1J_4+\epsilon_3J_3\right)P_{2}\left( 0 \right)\right]\nonumber\\
&&+\frac{ f_0}{m\varpi}\left[\left\{\Omega J_4-\frac{2\varpi}{\Omega-2\varpi}+\frac{\Omega}{2\varpi}\left\{\left(1+J_1\right)\epsilon_3+\left(1+J_2\right)\epsilon_1\right\}\right\}P_{1}\left( 0 \right)\right. \\ \nonumber\\
&&\left. -\frac{\Omega}{2\varpi}\left\{\left(1+J_2\right)\epsilon_3-\left(1+J_1\right)\epsilon_1\right\}P_{2}\left( 0 \right)\right]\nonumber\\
&&-\frac{\varpi}{m\Omega\left(\Omega-2\varpi\right)}\left[\epsilon_3P_{1}\left( 0 \right)+\epsilon_1P_{2}\left( 0 \right)\right]
\label{x1linear}
\end{eqnarray}
\begin{eqnarray}
\langle X_{2}\left(t\right)\rangle & = & \left[\left(1- \frac{1+J_2}{4}\right)X_{2}\left( 0 \right)+\frac{\left(1+J_1\right)}{4}\frac{P_{2}\left( 0 \right) }{m\varpi}\right]-\left[\left(\frac{1+J_1}{4}\right)X_{1}\left( 0 \right)+\frac{\left(1+J_2\right)}{4}\frac{P_{1}\left( 0 \right) }{m\varpi}\right]\nonumber\\
&& + \frac{\Omega f_0}{2}\left[J_4X_{1}\left( 0 \right)+J_3X_{2}\left( 0 \right)\right]-\frac{\varpi\left(\Omega-\varpi\right)}{\Omega\left(\Omega-2\varpi\right)}\left[\epsilon_3X_{1}\left( 0 \right)+\epsilon_1X_{2}\left( 0 \right)\right] \nonumber\\
&& +\frac{\left(\Omega-\varpi\right)}{2}\left[-\left(\epsilon_3J_3+\epsilon_1J_4\right)X_{1}\left( 0 \right)+\left(\epsilon_3J_4-\epsilon_1J_3\right)X_{2}\left( 0 \right)\right]\nonumber\\
&&+\frac{\Omega f_0}{4\varpi}\left[\left\{\left(1+J_1\right)\epsilon_3-\left(1+J_2\right)\epsilon_1\right\}X_{1}\left( 0 \right)+\left\{\left(1+J_2\right)\epsilon_3+\left(1+J_1\right)\epsilon_1\right\}X_{2}\left( 0 \right)\right]\nonumber\\
&&+\frac{1}{2m}\left[\left(\epsilon_1J_4+\epsilon_3J_3\right)P_{1}\left( 0 \right)-\left(\epsilon_3J_4-\epsilon_1J_3\right)P_{2}\left( 0 \right)\right]\nonumber\\
&&+\frac{ f_0}{m\varpi}\left[-\frac{\Omega}{2\varpi}\left\{\left(1+J_2\right)\epsilon_3-\left(1+J_1\right)\epsilon_1\right\}P_{1}\left( 0 \right)\right. \nonumber\\
&&\left. +\left\{\Omega J_4-\frac{2\varpi}{\Omega-2\varpi}+\frac{\Omega}{2\varpi}\left\{\left(1+J_1\right)\epsilon_3+\left(1+J_2\right)\epsilon_1\right\}\right\}P_{2}\left( 0 \right)\right]\nonumber\\
&&-\frac{\varpi}{m\Omega\left(\Omega-2\varpi\right)}\left[\epsilon_1P_{1}\left( 0 \right)-\epsilon_3P_{2}\left( 0 \right)\right]
\label{x1linear}
\end{eqnarray}
Similar expressions for the momentum expection values can also be written following the same algorithm. Once these phase space variables are in place they can be combined to construct the time evolution of all other dynamical quantities. This completes our solution for the interaction of a circularly polarised gravitational wave with a Landau system, i.e., charged particle, in the presence of a uniform magnetic background. Though the expressions are quite complicated two features of the solution are immediate noticeable. 
\begin{enumerate}
\item
First is the presence of three types of oscillatory terms, one intrinsic to the Landau system, with twice the cyclotron frequency $2\varpi$, indicated by $J_1$ and $J_2$ and the other two, indicated by $J_3$ and $J_4$, related to the GW; amongst them one exclusive to the GW and the other showing a coupling between the Landau system and the GW. We expect this last type to lead to interesting observational aspects if indeed a Landau system interacting with circularly polarized GW can be monitored.

\item
Second is the resonance feature near $\Omega=2\varpi$. This effect has been obtained earlier in a classical treatement \cite{emgw_classical} where one starts with the geodesic equation of the charged particle, instead of the geodesic deviation equation used in this paper. It is indeed reassuring that the condition of resonance obtained through our quantum mechanical analysis is in complete conformity with the classical analysis \cite{emgw_classical}. Also note that the specific choice of coordinate axes along the natural directional trio, defined by the polarization vectors of the system, although changes the computation/result of the present paper from that of our earlier work \cite{emgw_1_lin} where we considered only linearly polarized GW, the resonance point obtained in both cases are identical. Thus the resonance behaviour obtained by quantum mechanical treatement match with the classical result regardless of the linearly or circularly polarized nature of the GW.
\end{enumerate}

We would like to conclude with the observation that for a GW (with linear or circular polarization) interacting with Landau system which has an intrinsic oscillatory nature, one would essentially obtain a resonance behaviour. Hence, one can in principle detect the GW and find its frequency $\Omega$ by tuning the cyclotron frequency $\varpi$ which in turn depends on the applied magnetic field $B$ till resonance is achieved. A more realistic scenario can however be obtained if various forms of periodic GW signals with more than one frequency are used to carry out similar computations. We hope to report this in future.

\section*{Acknowledgement}
AS acknowledges the support by DST SERB under Grant No. SR/FTP/PS-208/2012.

\end{document}